# Statistical properties of DNA sequences revisited: the role of inverse bilateral symmetry in bacterial chromosomes

Marco V. José[1,*], Tzipe Govezensky[1], Juan R. Bobadilla[1]

[1]*Theoretical Biology Group, Instituto de Investigaciones Biomédicas, UNAM, Ciudad Universitaria, México D.F. C.P. 04510, México*

## Abstract

Herein it is shown that in order to study the statistical properties of DNA sequences in bacterial chromosomes it suffices to consider only one half of the chromosome because they are similar to its corresponding complementary sequence in the other half. This is due to the inverse bilateral symmetry of bacterial chromosomes. Contrary to the classical result that DNA coding regions of bacterial genomes are purely uncorrelated random sequences, here it is shown, via a renormalization group approach, that DNA random fluctuations of single bases are modulated by log-periodic variations. Distance series of triplets display long-range correlations in each half of the intact chromosome and in intronless protein-coding sequences, or both long-range correlations and log-periodic modulations along the whole chromosome. Hence scaling analyses of distance series of DNA sequences have to consider the functional units of bacterial chromosomes.

___________

*Corresponding author. Tel/Fax: 52-555-6223894;
*E-mail*: marcojose@biomedicas.unam.mx
[1]M.V.J. was financially supported by PAPIIT IN205702, UNAM, México. The authors thank Alma V. Lara and Alejandro Flores for computer programming assistance.

# 1. Introduction

A central issue of the statistical properties of DNA sequences was the finding of long-range correlations in intron-containing genes and in nontranscribed regulatory DNA sequences, but not in intronless genes [1]. Given the higher rate at which natural selection accepts different kinds of mutations in noncoding DNA, as compared to coding genome regions [2], the presence of long-range correlations in the former and weakly correlations in the latter demands for an explanation. However, the literature is inconclusive since a panoply of results can be found: the lack of power-law correlations in prokaryotes [2-5], the existence of power-law correlations covering all sizes in all prokaryotes [6] or the observation that the power spectrum of bacterial sequences flattens off in the low frequency limit thereby concluding that the fractal power-law behaviour of bacteria DNA does not always prevail throughout the entire DNA molecule [7-9]. Exponential distributions long-ranged and not power-law have been obtained for bacterial genomes [10]. By applying wavelet analysis, it was possible to discriminate between the fluctuations of single bases in noncoding regions (which behave like fractional Brownian motions), from protein-coding regions which cannot be distinguished from uncorrelated random Brownian walks [11]. Additionally, it has been contended that long-range correlations can be trivially caused by simple variations in nucleotide composition along DNA sequences [12]. Other studies have pointed out that both intron-containing and intronless DNA sequences exhibit long-range correlations [13-15]. Therefore it is still an open question whether the long-range correlation properties are different for intronless and intron-containing coding regions.

The above controversies primarily arise due to different methods or viewpoints used in the analysis. Furthermore, most of the previous analyses do not consider the whole structural architecture of the bacterial chromosomes. These analyses were made without considering replication or recognizable functional units.
In a previous work cumulative position plots (CPPs) of all 64 DNA tri-nucleotides (triplets) along the *B. burgdorferi* chromosome were obtained and they revealed high correlations in complementary triplet frequencies (parity) between opposing leading and lagging strands (chromosomal halves) [17]. Parity was also observed for di- and mononucleotides and, to a limited extent, for higher order n-tuples. Using *ad hoc* shuffled sequences it was demonstrated that the cumulative distributions were not merely the consequence of base composition. Because parity occurred between opposing chromosomal halves it was well established that bacterial chromosomes posses an inverse bilateral symmetry (IBS), that seems to be a universal type of symmetry in Eubacteria [17].
In this work, the hypothesis that the IBS observed in whole bacterial chromosomes constitutes an organizing principle for a better understanding of the statistical properties of DNA sequences is put forward.

Herein we focus our attention to the correlation properties of the whole chromosome and of exclusively protein-coding regions by considering the functional units of the chromosome such as the replichores that correspond to the chromosomal halves of the bacterial genomes (see below). Bacterial chromosomes when compared with Eukaryotes, which have several replication origins, offer this possibility of analysis since the location of the origin of replication and terminus sites are well known, and leading and lagging strands can easily be distinguished. Further bacterial chromosomes are mainly composed of coding regions. The bacteria *Borrelia burgdorferi* was chosen because of the small size (910 kb) of its linear chromosome [16] but the result can be extended for several other Eubacteria.

Instead of using the classical stochastic random walk mapping rules of DNA (e.g. the purine-pyrimidine (RY) rule), the distance series of any n-tuple (or any DNA sequence of finite arbitrary length) along any segment of the chromosome (e.g. chromosomal halves) were determined.

For a given n-tuple its actual position along the whole chromosome was determined and from this either a CPP or the actual distance series (distance measured in bases) of that particular n-tuple were obtained. In other words, we can directly visualize how a given n-tuple is distributed along the whole chromosome. To our knowledge, no studies have been made by analyzing the actual distance series of the distribution of n-tuples along functional units of bacterial chromosomes.

The distance series of single bases and triplets and their corresponding complementary bases or triplets were analysed by means of detrending fluctuation analysis (DFA) and via a renormalization group (RG) approach from which the Hurst exponent and hence the fractal dimension can be estimated. Both the RG and the DFA techniques rendered exactly the same results.

The analysis below focuses on the variability of scales of distance series of a given n-tuple taking as a unit of study both the whole chromosome and each half of it. Herein it is proved that when the distance series of single bases or triplets within each half of the intron-containing bacterial chromosome are analysed they result in random monofractals with log-periodic modulations or in monofractal series with long-range correlations, respectively. The magnitude of the scaling exponent of a single base or a triplet is the same as that of its corresponding complementary base or triplet in the other half of the chromosome according to the IBS of the whole chromosome. When the whole bacterial chromosome is considered the distance series of triplets are nonstationary due to the constraints imposed by the IBS, and then multiscale series are obtained. For the whole chromosome the variability of scales reflected a multiplicity of scales rather than a single parameter, which can be accounted for with a more general renormalization scaling model. This model predicts a dominant power-law decrease with the size of the window modulated by a log-periodic function.

Similarly, in exclusively protein-coding regions the scaling behaviours between chromosomal halves at the level of complementary bases, dinucleotides, and triplets are not identical to those for the whole intact chromosome, indicating that IBS is a source of a multiscale phenomenon occurring in codons, and not just in triplets. Similar values of long-ranged scaling exponents between chromosomal halves are obtained when all codons in each chromosomal half are compared altogether but they are different when codons in the whole leading and the whole lagging strands are analysed separately.

*1.1. Bacterial chromosomes*

Replication of the bacterial chromosomal DNA starts at the origin (*ori*) and the two replication forks (or replichores) proceed bi-directionally (in opposite directions) up to the terminus (*ter*) region [18]. The two replication forks need to reach the *ter* region at the same time to optimise the replication process. Evidence has been obtained that *ori* and *ter* should have a physical balance between them, i.e., opposite to each other, on the chromosome to guarantee a synchronous completion of the bi-directional chromosomal replication [19]. To a close approximation, *ori* and *ter* are equidistantly located in circular chromosomes. In analogy, in the linear chromosome of *B. burgdorferi* there is a single *ori* located in the middle of the sequence [20,21]. According to the replication mechanism, most bacterial chromosomes can be divided into two functional units or replichores [22] and these units are approximately equivalent to chromosomal halves. Since chromosomal halves often differ in base composition the alternative name of chirochores have been used for them [23]. During bi-directional duplication of the chromosome both complementary strands are simultaneously replicated in the 5' to 3' direction but while replication is continuous for one strand (leading strand) it is discontinuous for the other (lagging strand) [18]. Differences among the two strands extend to their coding potential and there are more open reading frames (ORF) and a larger proportion of highly expressed genes in the leading than in the lagging strand [24]. Codon usage is also different between the leading and lagging strands [25,26]. Disparities between leading and lagging strands have been ascribed to biased mutation and/or selection associated to transcription [27] or replication [24,28]. Discrepancies between the two strands have been considered elements of chromosomal asymmetry [29]. However, when whole bacterial chromosomes are considered they are not compositionally asymmetric but rather they display IBS [17].

## 2. Materials and Methods

*2.1. Data*

The complete sequence of *B. burgdorferi* was retrieved from the NCBI, Genbank resource from the NIH (http://www.ncbi.nlm.nih.gov) with the following accession number: NC_001318.
By means of an *ad hoc* program using Matlab version 5.3 the actual position of any n-tuple along the whole chromosome was determined and from this either a CPP or the actual distance series of that particular n-tuple were obtained. Randomisation of sequences (shuffling) was done by an *ad hoc* program written in language C. This program was also useful for translating the DNA sequence into aminoacids or for backtranslation to codons and it was also used for triplet or codon shuffling. To perform the present analyses, DNA sequences in their single stranded form were divided at the known or proposed *ori* and *ter* [21,25,28] so as to produce two segments of equal length (chromosomal halves).

The following intronless sequences (neither intergenic regions nor ribosomal or transfer RNAs) for *B. burgdorferi* were assembled *in silico*:

1. Open reading frames (ORF) in their coding-wise orientation (ORF-CW) retrieved from NCBI as *.ffn files. This sequence contains all ORF joined in-frame one after the other, including the stop codons.
2. ORF in their original order and orientation (ORF-OOO). Such ORF-OOO sequence was created by joining all reported ORF one after the other as naturally ordered and oriented in the chromosome.

The following randomised sequences were also prepared:
1. Shuffled whole chromosome as bases (SWB)
2. Shuffled whole chromosome as triplets (SWT)
3. Shuffled chromosomal halves as bases (SHB)
4. Shuffled chromosomal halves as triplets (SHT)

The computer programs for performing the present analyses are available upon request.

## 2.2. Detrended Fluctuation Analysis

The DFA method enables the detection of long-range correlations embedded in a patchy landscape, also avoiding the spurious detection of apparent long-range correlations that are an artefact of patchiness. The DFA method comprises the following steps [2]: The interval distance series (of total length $N$) is first integrated, $y(k) = \sum_{i=1}^{k}[B(i) - Bave]$, where $B(i)$ is the *ith*-distance interval and $B_{ave}$ is the average distance interval. Next the integrated distance series is divided into boxes of equal length, $\ell$. In each box of length $\ell$, a least-squares line is fitted to the data (representing the trend in that box). The $y$-ordinate of the straight line segments is denoted by $y_\ell(k)$. Next the integrated distance series, $y(k)$, is detrended by substracting the local trend, $y_\ell(k)$, in each box. The root mean square (r.m.s.) fluctuation of this integrated and detrended distance series is calculated by:

$$F(\ell) = \sqrt{\frac{1}{N}\sum_{k=1}^{N}\left[y(k) - y_\ell(k)\right]^2} \ . \tag{1}$$

This computation is repeated over all distance-scales (box sizes) to provide a relationship between $F(\ell)$, the average fluctuation as function of box size, and the box size $\ell$ (i.e. the size of the window of observation is the number of a given n-tuple in one box). Typically, $F(\ell)$ will increase with box size $\ell$. A linear relationship on a double log graph indicates the presence of scaling. Under such conditions, the fluctuations can be characterized by a scaling exponent $\alpha_{DFA}$, the slope of the line relating $\log F(\ell)$ to $\log \ell$.

If the data is uncorrelated, the integrated value, $y(k)$, corresponds to random walk, and therefore $\alpha_{DFA} = 0.5$ [31]. If there are only short-term correlations, the initial slope may be different from 0.5, but $\alpha_{DFA}$ will approach 0.5 for large window sizes. An $\alpha_{DFA}$ greater than 0.5 and less than or equal to 1.0 indicates persistent long-range power-law correlations such that a large (compared to the average) incidence interval is more likely to be followed by large interval and vice versa. In contrast, $0 < \alpha_{DFA} < 0.5$ indicates a different type of power-law correlation such that large and small values of the time series are more likely to alternate. A special case of $\alpha_{DFA} = 1$ corresponds to $1/f$ noise [2,31]. For $\alpha_{DFA} \geq 1$, correlations exist but cease to be of a power-law form.

## 2.3. Renormalization Group: Hurst exponent and Fractal Dimension

Let us examine how the relative dispersion $(RD)$ changes as a function of the number of adjacent data elements we aggregate. To see this we aggregate $n-$ adjacent data points, so that the $j-$th element in such an aggregation is given by: $Y_j^{(n)} = Y_{nj} + Y_{nj-1} + Y_{nj-2} + ... + Y_{nj-(n-1)}$. In terms of these new data the average is defined as the sum over the total number of data points, where the bracket denotes the closest integer value, and is the original number of data points, i.e.,

$$\overline{Y}^{(n)} = \frac{1}{[N/n]} \sum_{j=1}^{[N/n]} Y_j^{(n)} Y_j^{(n)} = n\overline{Y}^{(1)}. \qquad (2)$$

The variance for a monofractal time series is similarly given by [32]:

$$Var(\overline{Y}^{(n)}) = n^{2H} Var(\overline{Y}^{(1)}), \qquad (3)$$

where $H$ is the Hurst exponent, and the superscript $(1)$ on the average variable indicates that it was determined using all the original data without aggregation and the superscript $(n)$ on the average variable indicates that it was determined using the aggregation of $n$ data points. Thus, the $RD$ for an aggregated data set is:

$$RD^{(n)} = \frac{\sqrt{Var(\overline{Y}^{(n)})}}{\overline{Y}^{(n)}} = \frac{\sqrt{n^{2H}Var(\overline{Y}^{(1)})}}{n\overline{Y}^{(1)}} = n^{H-1}RD^{(1)}, \qquad (4)$$

which is precisely an inverse power-law (IPL) in the aggregation number for the Hurst exponent in the interval $0 \leq H \leq 1$. It is well established [32,33], that the exponent in such scaling equations is related to the fractal dimension, $D$, of the underlying distance series by $D = 2 - H$. A simple monofractal time series, therefore, satisfies the IPL relation for the $RD$ given by Equation (4) which can be expressed by the linear regression relation [34]:

$$\ln RD^{(n)} = \ln c + m \ln n, \qquad (5)$$

where $m = 1 - D$. The significance of the fractal dimension can be determined using the autocorrelation function, which for nearest neighbour data points can be written as [34,35]:

$$r_1 = 2^{3-2D} - 1. \qquad (6)$$

If the fractal dimension is given by $D = 1.5$, the exponent in Equation (6) is zero and consequently $r_1 = 0$, indicating that the data points are linearly independent of one another. This would be the case for a slope of $m = -0.5$ in the log-log plots of the $RD$ versus aggregation number, $n$. This fractal dimension corresponds to an uncorrelated random process with normal statistics, often referred to as Brownian motion. If, on the other hand, the nearest neighbors are perfectly correlated, $r_1 = 1$, the irregularities in the distance series are uniform at all distances and the fractal dimension is determined by the exponent in Equation (6) to be $D = 1$. A fractal dimension of unity implies that the distance series is regular, such as it would be for simple periodic motion. Most time or distance series have fractal dimensions that fall somewhere between the two extremes of Brownian motion with $D = 1.5$, and complete regularity with $D = 1$.

*2.4. Multiplicative Preferential Scale*

The deviation of the aggregated relative dispersion data from the best IPL fit, implies that an underlying mechanism exists that is producing this effect. In order to incorporate the variability of distances for a given n-tuple along the chromosome it is necessary to take cognisance of the fact that rather than a single scaling parameter, there are a number of such parameters each with a different frequency of occurrence.

Consider a random function $f(n,\lambda)$ that is related to the theoretical relative dispersion at aggregation level $n$, but it is characterized by a distance scale $\lambda$. It is known that the $RD$ does not have a characteristic distance scale, in fact, at each level $n$, there are a number of such scales present. The fluctuations in $f(n,\lambda)$ can then be characterized by a distribution of $\lambda's$. Lets denote the probability that a particular scale is present in the relative dispersion in the interval $(\lambda, \lambda + d\lambda)$ by $p(\lambda)d\lambda$. The theoretical relative dispersion at the *nth* aggregation level is then:

$$\overline{RD}^{(n)} = \int_0^\infty f(n,\lambda)p(\lambda)d\lambda, \qquad (7)$$

where the dependence on time or distance scales are averaged out. If the stochastic process is sharply peaked with only a single scale, say $\lambda = \overline{\lambda}$, then $p(\lambda) = \delta(\lambda - \overline{\lambda})$ and Equation (7) reduces to the exponential $RD(n,\lambda) = RD_0 e^{-\lambda n}$. However, if the distributions of scales contain finite central moments, then the theoretical relative dispersion given by Equation (7) will contain this scale.

Consider any distribution $p(\lambda)$ that has central finite moments, say a mean value $\overline{\lambda}$. Now following [Montroll and Shlesinger [31]], an amplification mechanism is applied such that $p(\lambda)$ has a new mean value $\overline{\lambda}/b$:

$$p(\lambda/\overline{\lambda}) \rightarrow p(b\lambda/\overline{\lambda}), \qquad (8)$$

and this amplification occurs with relative frequency $1/a$. This amplification is applied a second time so that the scaled mean is again scaled and the new mean is $\overline{\lambda}/b^2$ and occurs with relative frequency $1/a^2$. This amplification process is repeated over and over again and eventually generates the new unnormalized distribution:

$$P(\phi) = p(\phi) + \frac{1}{a} p(b\phi) + \frac{1}{a^2} p(b^2\phi) + ..., \qquad (9)$$

in terms of the dimension variable $\phi = \lambda/\overline{\lambda}$. The distribution (9) is now used to evaluate the observed $RD$ in terms of the theoretical relative dispersion:

$$RD^{(n)} = \overline{RD}^{(n)} + \frac{1}{a}\overline{RD}^{(n/b)} + \frac{1}{a^2}\overline{RD}^{(n/b^2)} + ... \qquad (10)$$

or neglecting higher order terms,

$$RD^{(n)} = \frac{1}{a}\overline{RD}^{(n/b)} + \overline{RD}^{(n)}. \qquad (11)$$

Equation (11) has the form of a renormalization group (RG) transformation. The exact solution can be separated into an analytical and a singular part, the latter part dominating. The singular part of the solution, denoted by the subscript *sing*, satisfies the functional equation:

$$RD^{(n)}_{sing} = \frac{1}{a} RD^{(n/b)}_{sing}. \qquad (12)$$

The solution to Equation (12) is known to be [36]:

$$RD^{(n)}_{sing} = \frac{A(n)}{n^{\alpha_{RG}}}, \qquad (13)$$

where by direct substitution we obtain for the power-law index of the RG,

$$\alpha_{RG} = \frac{\ln a}{\ln b}, \qquad (14)$$

and the modulation function:

$$A(n) = A(n/b) = \sum_{k=-\infty}^{\infty} A_k e^{-ik2\pi \frac{\ln n}{\ln b}}.  \qquad (15)$$

Thus, via a RG approach the aggregated *RD* displays a dominant IPL with an index given by (14) and is modulated by a function that varies logarithmically with a fundamental period *ln b*. Note that the solution (13) does not depend on the specific form of the distribution of scales, only on the condition that the distribution has a finite mean. This mean is then leveraged by means of the amplification mechanism to yield the modulated IPL. If the function $A(n)$ is constant then (13) reduces to (5) and the underlying phenomenon becomes monofractal.

The goal is to determine if the distance series of n-tuples (particularly mono-, di-, and, triplets) in complete prokaryote genomes obey an IPL with or without a log-periodic modulation. By using Equation (13) the distance series can be fitted to a model with RG properties. Here the following fitting function is used [34,36]:

$$\ln RD^{(n)} = \ln a_1 + a_2 \ln n + a_3 \cos[a_4 \ln n], \qquad (16)$$

where the four parameters are obtained by fitting the aggregated *RD* plotted versus the aggregation number, $n$, on a log-log graph. Note that the modulation term in (16) is a simplification of the general function $A(n)$ expressed in Equation (15). The Equation (16) is related to Equation (13) if we associate the parameters as follows: $a_1 = a$; $a_2 = -\alpha_{RG}$ and $a_4 = 2\pi / \log b$. In practice, the slope and intercept obtained by the linear regression of Equation (5) are used as seed values in order to perform, via the Levenberg-Marquardt procedure, the nonlinear fitting of Equation (16).

All the figures related to the RG operation were obtained by aggregating one by one the time series and the length of the final time series is equal to 100 except in the case of single bases in which the length of the final time series is equal to 1000.

## 3. Results

### 3.1. *Scaling properties of single bases*

The CPPs for A (adenine) and T (thymine) in the complete genome of *B. burgdorferi* form a rhombus as can be seen in Fig. 1 (note that for convenience the dependent variable of the CPPs is in the abscissa whilst the independent variable is in the ordinate). This means that IBS is reflected at the level of a single base and its complementary. Each single base and its complementary form rhombi whose slopes and lengths of opposite sides, i.e., opposing leading and lagging strands, are practically equal, and both midpoints of the rhombi coincide with the *ori* site (~ 4,575 kb). Note in the abscissa that the total number of A's is approximately equal to the number of T's for the whole chromosome as predicted by the Chargaff's second parity rule [37].

There are base disparities in the transition between leading and lagging strands in chromosomal halves, and these transitions have been dubbed GC-skews since they are usually recognized by shifts in G (guanine)+C (cytosine) content [23]

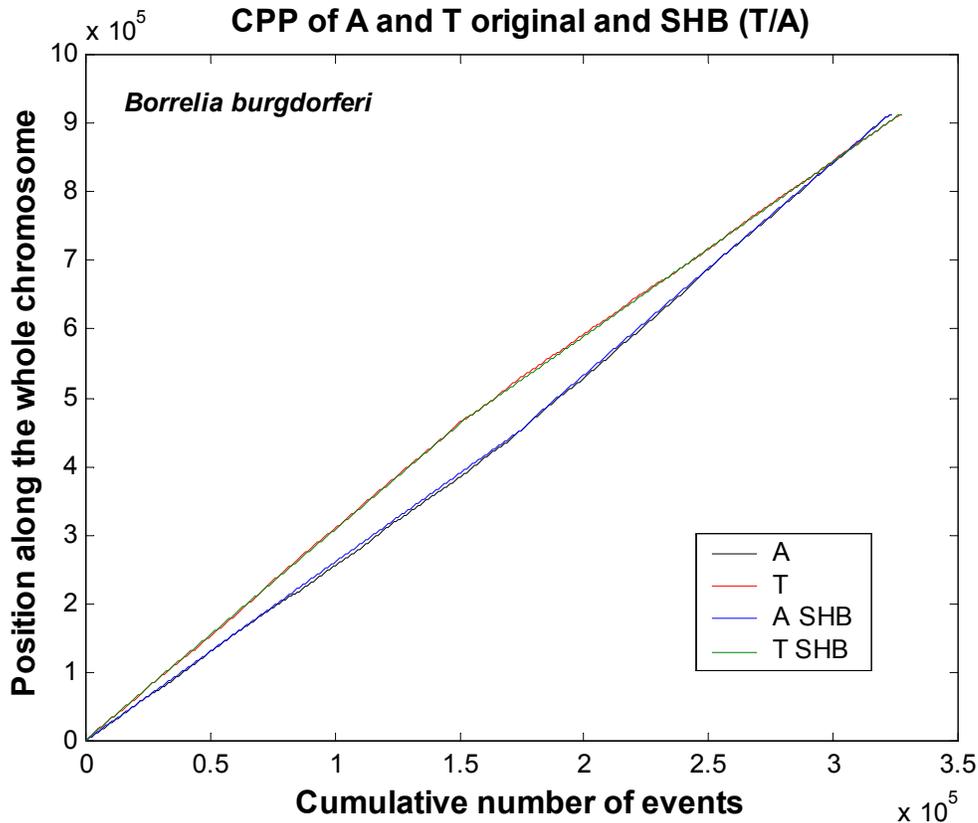

**Fig. 1:** Cumulative position plots for A and T in a *B. burdorferi* sequence either in its intact form or after shuffling the chromosomal halves at the level of bases (SHB). Note that axes are not used in the standard way and the dependent variable (cumulative position value) is in the abscissa whilst the independent variable (nucleotide position) is in the ordinate. The designation T/A indicates that distributions for the monoplet on the left side (i.e. T) correspond to the two left sides of the rhombi, while distributions for the other monoplet (i.e. A) correspond to the right sides of rhombi;

To explore how important the natural order of a single base was for IBS, chromosomal halves were shuffled independently and at the level of bases (SHB). The CPPs for A and T from the SHB control sequences generated also a rhombus that was visually indistinguishable from the actual biological rhombus (see Fig. 1). However, the aggregation analysis for the original sequences of both A in the 1$^{st}$ half and T in the 2$^{nd}$ half of the chromosome showed a similar linear decay of the $RD$ as the size of the window $n$ increased in a log-log plot, but there is a departure from this linear decay at larger window sizes (see Fig. 2). The fitting of the more general renormalization scaling model (Equation (16)) showed that the decay is still of the random type but it is modulated by a log-periodic function. As expected from IBS the scaling parameter for A in the 1$^{st}$ half ($\hat{\alpha}_{RG} = 0.49$) is very similar to the one obtained for T in the 2$^{nd}$ half ($\hat{\alpha}_{RG} = 0.48$) of the chromosome. The decay of $RD$ for both A and T is similar to the decay observed from the corresponding shuffled controls of A and T up to the region in which the window size lies between 100 and 1000, although the amplitude (y-intercept) is different between the biological ones and the shuffled controls. The scaling behaviour for both A and T obtained from their corresponding SHB sequences showed straight lines (monofractal pattern) with negative slopes of the order of 0.5, clearly indicating a random behaviour for all window sizes. Note however, that the estimates of $\hat{\alpha}_{RG}$ for A and T, 0.49 and 0.48, respectively, are very similar to the scaling values obtained for A and T from the SHB sequences (0.5 for A in the 1$^{st}$ half and 0.51 for T in the 2$^{nd}$ half). Therefore, the scaling properties of A and T in each half of the chromosome exhibit a random behaviour but slightly edited by a log-periodic function particularly for large window sizes. The same results were obtained for the nucleotides G (guanine) and C (cytosine) (not shown).

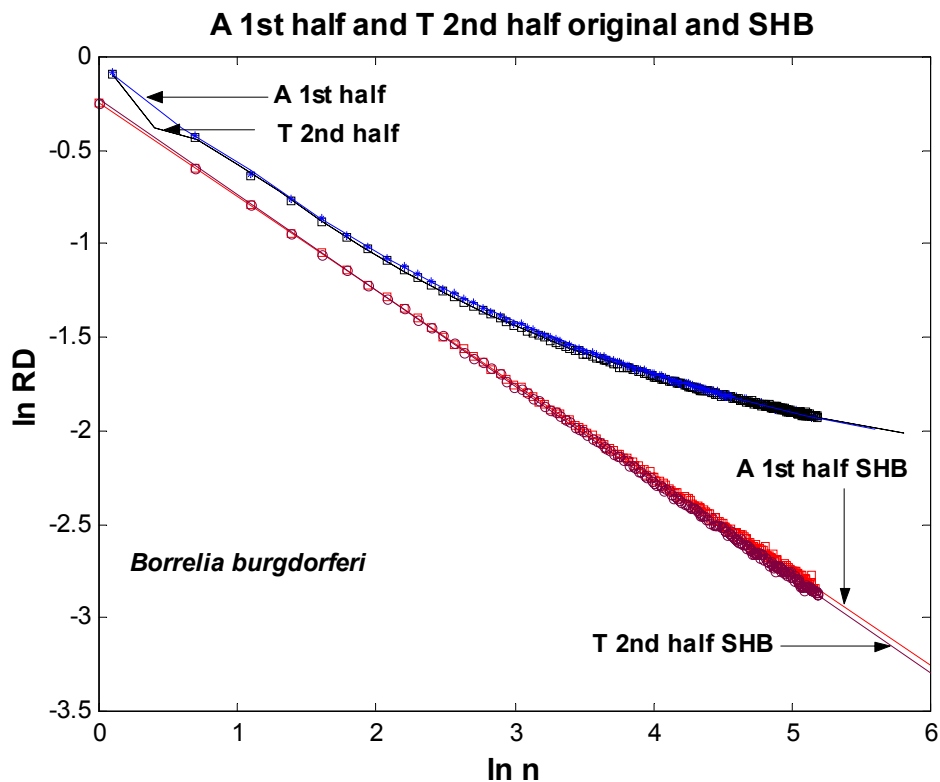

**FIG. 2:** $\ln RD$ versus $\ln n$ of A in the 1st half (fitted equation: $RD = \frac{-0.11}{n^{0.49}} e^{0.33\cos[1.36\ln(n)]}$) and T in the 2nd half (fitted equation: $RD = \frac{-0.11}{n^{0.48}} e^{0.31\cos[1.48\ln(n)]}$) of the chromosome. For A in the 1$^{st}$ half of the chromosome from the SHB sequence the fitted equation is: $\ln RD = -0.5\ln(n) - 0.24$ ($r^2 = 0.99$) and for T in the 2$^{nd}$ half of the chromosome the fitted equation is: $\ln RD = -0.51\ln(n) - 0.22$ ($r^2 = 0.99$); **(c)** DFA of A (diamonds) and T (crosses) for the first 100 kbp of the *B. burgdorferi* chromosome; the corresponding shuffled controls are for A (asterisks) and for T (pluses).

To verify with a second independent test the observed behaviour with the aggregation analysis, the DFA technique was used in order to see if the fluctuations at distances larger than 100 bases could indeed be correlated. In Fig. 3 the results of the DFA to both distance series of A (crosses) and T (pluses) from the first 100 kb of the original DNA sequence of *B. burgdorferi*, as well as the resulting DFA of single realizations of the corresponding shuffled sequences for A (diamonds) and T (asterisks) are shown. A good linear fit of the $\log F(\ell)$ versus $\log \ell$ should be proportional to $\ell^{\alpha_{DFA}}$, where $\alpha_{DFA}$ is the single exponent describing the correlation properties of the entire range of distance-scales, and the parameter $y_{int}$ is the y-intercept representing the amplitude of the r.m.s. of the fluctuations. Note the similar values of the scaling exponents but different values of the amplitudes for A and T: for A original $y_{int} = -0.37$ and $\hat{\alpha}_{DFA} = 0.6$; for T original $y_{int} = -0.24$ and $\hat{\alpha}_{DFA} = 0.59$;

However, in some of the scales we found that the DFA plot was not strictly linear but rather consisted of at least three distinct linear regions of different slopes separated by two breakpoints, BP1 and BP2. This observation suggests that there is a short-range scaling exponent, $\alpha_{DFA1}$, over a period from $\ell = 0.5$ up to BP1=2.25 (approximately 178 bases), a long-range scaling exponent, $\alpha_{DFA2}$, from BP1 to BP2=3, from 178 bases to 1000 bases, and another long-range scaling exponent, $\alpha_{DFA3}$, from BP2 to 100 kb.

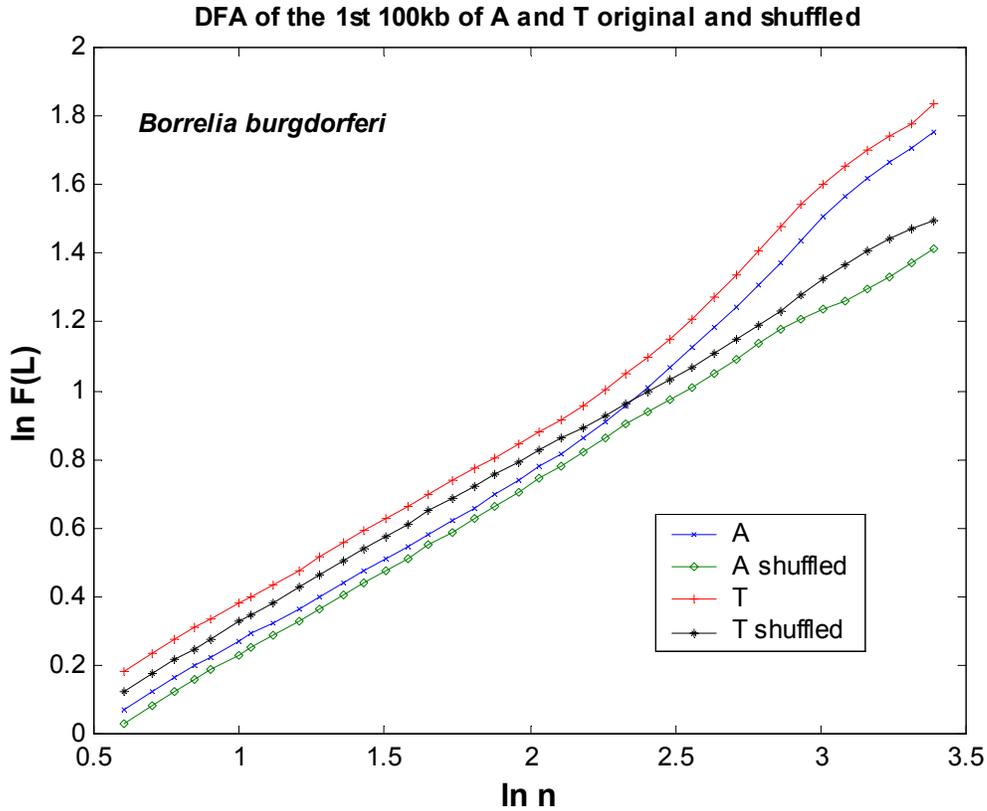

**FIG. 3:** DFA of A (diamonds) and T (crosses) for the first 100 kbp of *B. burgdorferi* chromosome; the corresponding shuffled controls are: A (asterisks) and T (pluses).

In brief, we note that for both distance series of A and T, DFA detects two crossovers, namely, within the first ~178 bases there is one behaviour where $\hat{\alpha}_{DFA1} \approx 0.5$ (uncorrelated random noise); from ~178 bases to 1 kb, a long-range scaling exponent $0.5 < \alpha_{DFA2} \ll 1$ is observed (which means that there are weakly persistent long-range correlations); and between 1 kb to 100 kb, $0 \ll \alpha_{DFA3} < 0.5$ (indicating weakly anti-persistent power-law correlations). To rule out the possibility that these three different scaling factors could be the result of an artefact note that the scaling exponent for the shuffled sequences of A and T both give an overall scaling exponent of $\hat{\alpha}_{DFA} \approx 0.5$ (for A shuffled $y_{int} = -0.27$ and $\hat{\alpha}_{DFA} = 0.49$ and for T shuffled $y_{int} = -0.17$ and $\hat{\alpha}_{DFA} = 0.5$). It is interesting to remark that the scaling exponents of the distance series for both A and T from the actual biological DNA sequence are of the order of 0.5 (uncorrelated) and they are practically indistinguishable from their corresponding shuffled controls up to 178 bases.

*3.2. Scaling properties of triplets along the chromosome*

When the distance series of the triplets ATG and CAT are analysed along the whole bacterial chromosome, the aggregation analysis do not give straight lines as it is illustrated in Fig. 4. Both curves are fitted to Equation (16) indicating that they follow an IPL with log-periodic modulation, i.e., they are multifractals. In addition to the IPL of the average distance on $RD$ there appears to be a log-periodic variation of the data about this power-law behaviour. For the ATG triplet (asterisks) the nonlinear statistical fitting gives the following equation:

$$RD^{(n)} = \frac{0.35}{n^{0.37}} e^{0.74\cos[0.53\ln n]}$$

whereas for the triplet CAT (pluses) the resulting fitted equation gives:

$$RD^{(n)} = \frac{0.27}{n^{0.38}} e^{1.02\cos[0.42\ln n]}$$

Note the similar values of the estimates of the scaling exponents for ATG ($\hat{\alpha}_{RG} = 0.37$) and CAT ($\hat{\alpha}_{RG} = 0.38$) that indicate long-range correlations.

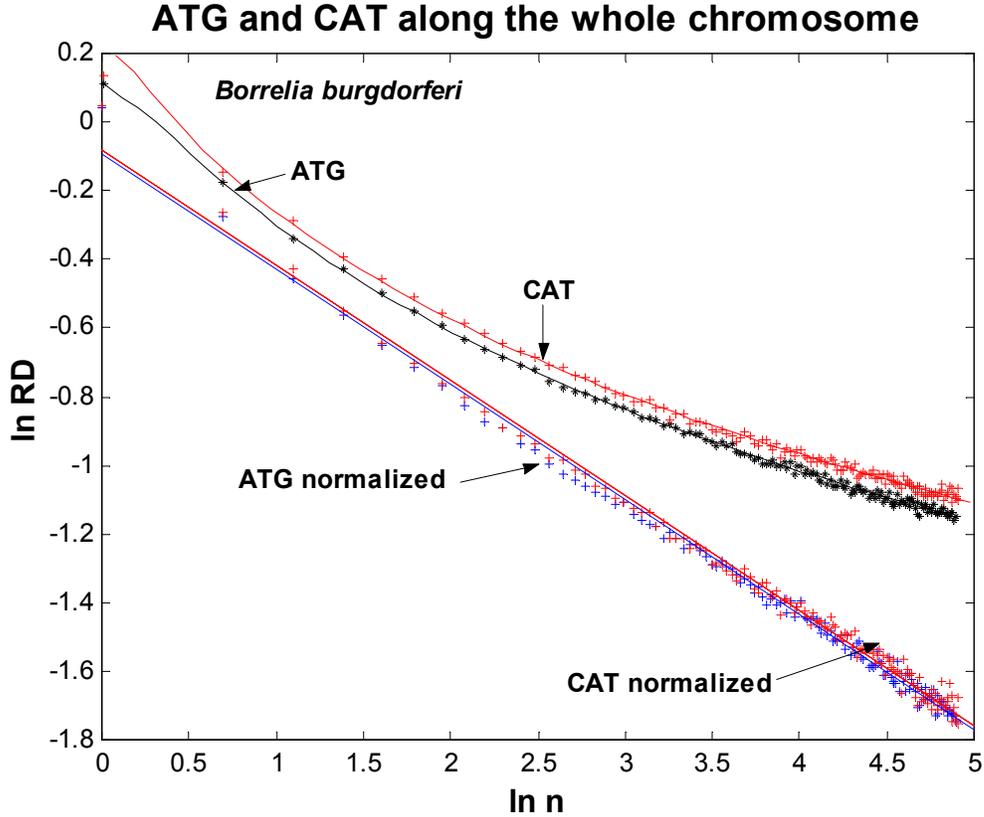

**FIG. 4:** Aggregation analysis of the triplets ATG and CAT along the whole chromosome. The corresponding fitted equations are given in the text. The fitted equation for the normalised distance series of ATG and CAT are, respectively, $\ln RD = -0.33 \ln n - 0.11$ ($r^2 = 0.99$) and $\ln RD = -0.33 \ln n - 0.09$ ($r^2 = 0.99$)

If the distance series of each triplet in each half of the chromosome is normalized by its respective mean distance, the secular trend is removed and the whole series now displays a long-ranged monofractal behaviour as can be seen in the figure for both triplets ATG and CAT. In this case the slopes of these straight lines are related to the Hurst exponent by means of Equation: $H = 1 + m$, so that $H = 0.67$ for both ATG and CAT. By this simple normalization we do not break the IBS of the bacterial chromosome but now the values of the scaling exponents for both ATG and CAT are corrected from those obtained when the whole chromosome is taken into account. We remark that what is biologically meaningful is the fact that the fluctuations of a given n-tuple in one half of the chromosome are related to the fluctuations of the complementary n-tuple in the other half.

In Fig. 5 the results of the aggregation analysis of the triplet ATG for both halves of the chromosome are illustrated. Straight lines arise indicating that the corresponding distance series are monofractals: for the 1st half of the chromosome the slope of the aggregation analysis gives $H = 0.67$ and, for the 2nd half $H = 0.64$. Using the relation, $D = 1 - m$, the corresponding fractal dimension of ATG for the 1st half is $D=1.32$ and for the 2nd half $D=1.36$. These estimates are in turn similar to the fractal dimensions obtained for the triplet CAT in the 2nd ($D=1.32$) and in the 1st half ($D=1.35$) of the chromosome, respectively (see caption of Fig. 5).

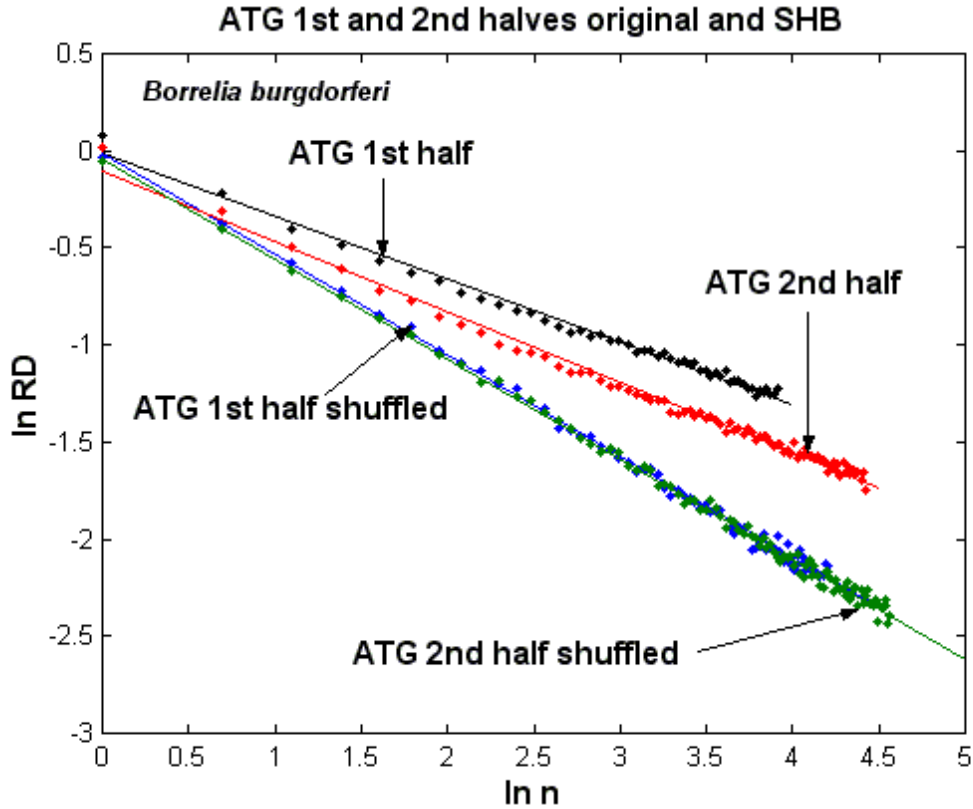

**FIG. 5:** Aggregation analysis of the triplet ATG by chromosomal halves. For the 1st half the fitted equation is: $\ln RD = -0.32 \ln n - 0.01$ ($r^2 = 0.99$); for the 2nd half the fitted equation is: $\ln RD = -0.36 \ln n - 0.1$ ($r^2 = 0.99$). The fitted equations for the corresponding shuffled controls at the level of bases are for the 1st and 2nd halves, respectively: $\ln RD = -0.52 \ln n - 0.01$ ($r^2 = 0.99$) and $\ln RD = -0.51 \ln n - 0.04$ ($r^2 = 0.99$); $\ln RD$ versus $\ln n$ of the triplet CAT by chromosomal halves. For the 1st half the fitted equation is: $\ln RD = -0.36 \ln n - 0.1$ ($r^2 = 0.99$); for the 2nd half the fitted equation is: $\ln RD = -0.33 \ln n - 0.02$ ($r^2 = 0.99$). The fitted equations for the corresponding shuffled controls at the level of bases are for the 1st and 2nd halves, respectively: $\ln RD = -0.5 \ln n - 0.08$ ($r^2 = 0.99$) and $\ln RD = -0.51 \ln n - 0.04$ ($r^2 = 0.99$) (not shown).

In Fig. 6 the DFA of the distance series of the triplets ATG and CAT for each half of the bacterial chromosome are presented. Note that the slope of the DFA of ATG in the 1st half ($\hat{\alpha}_{DFA} = 0.68$) is similar to the slope of the DFA of CAT in the 2nd half ($\hat{\alpha}_{DFA} = 0.67$), and that the slope of the DFA of ATG in the 2nd half ($\hat{\alpha}_{DFA} = 0.63$) is similar to that of CAT in the 1st half ($\hat{\alpha}_{DFA} = 0.64$). The estimates of the Hurst exponents obtained via the aggregation analysis are in turn equal to the slopes of the corresponding DFA by halves of the chromosome. The slopes obtained by using DFA of any triplet and its complementary are practically indistinguishable if the chromosomal halves are taken into account. These results are consistent with the IBS of the whole chromosome. Similarly, the amplitude of ATG in the 1$^{st}$ half ($y_{int} = 1.19$) is very similar to the amplitude of CAT in the 2$^{nd}$ half ($y_{int} = 1.2$), whereas the amplitude of ATG of the 2$^{nd}$ half ($y_{int} = 1$) is similar to that of CAT in the 1$^{st}$ half ($y_{int} = 0.98$). The slope of the corresponding shuffled controls by chromosomal halves are: ATG 1$^{st}$ half $\hat{\alpha}_{DFA} = 0.47$; CAT 2$^{nd}$ half $\hat{\alpha}_{DFA} = 0.49$; ATG 2$^{nd}$ half $\hat{\alpha}_{DFA} = 0.48$; CAT 1$^{st}$ half $\hat{\alpha}_{DFA} = 0.49$ (not shown).

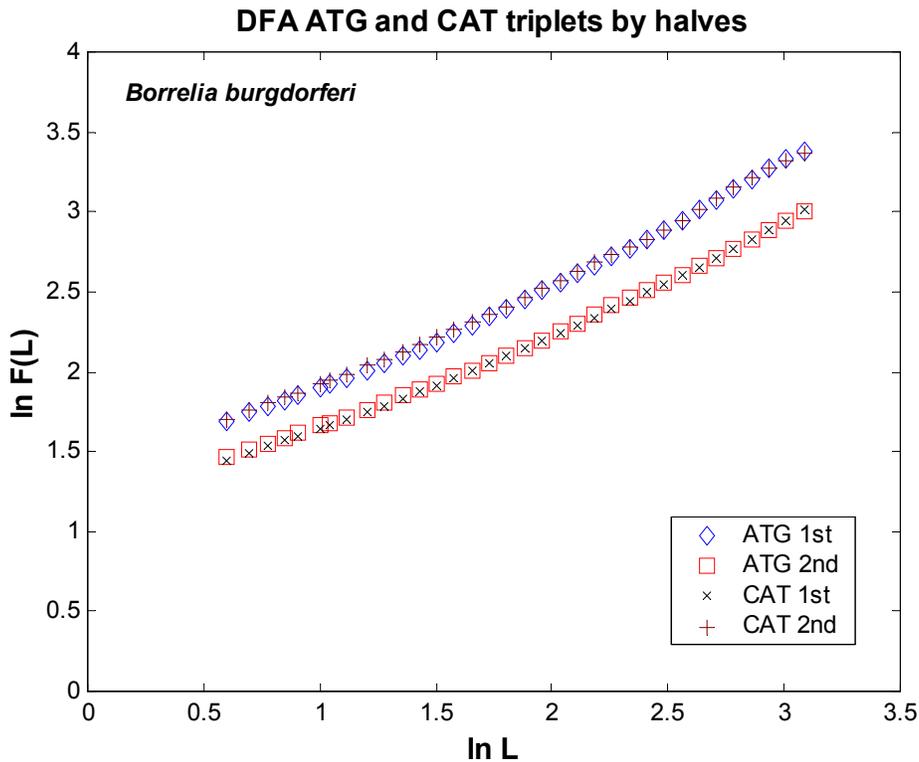

**FIG. 6:** DFA of the triplets ATG and CAT by chromosomal halves of the original chromosome of *B. burgdorferi*.

### 3.3. Scaling properties of triplets along protein-coding regions

In this section it is shown that whilst all distance series of triplets along protein-coding regions assembled in the ORF-CW sequence exhibit a linear decay with long-range correlations scaling exponents, they do not retain the IBS.

Fig. 7 shows that monofractal relations arise between the $\ln RD$ versus the $\ln n$ for the distance series of ATG and CAT in the ORF-CW sequence. In this case the behaviour of both ATG ($m=-0.39\ (H=0.6; D=1.39; r_1=0.16)$) and CAT ($m=-0.36\ (H=0.63; D=1.36; r_1=0.21)$) display long-range correlations, whereas their corresponding shuffled controls are random monofractals: for ATG: $m=-0.5\ (H=0.49; D=1.5; r_1=0)$ and, for CAT: $m=-0.49\ (H=0.5; D=1.49; r_1=0.01)$.

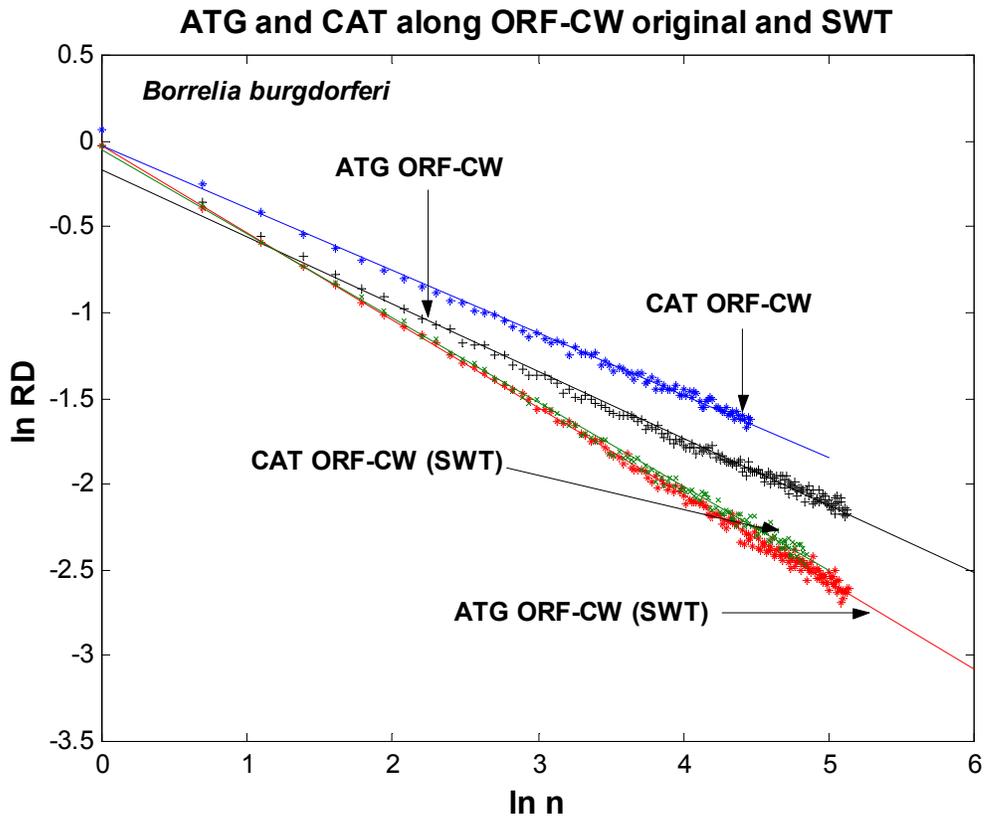

**FIG. 7:** Aggregation analysis of ATG (pluses) and CAT (asterisks) in the ORF-CW sequence and their corresponding SWB. The fitted equations for ATG and CAT are, respectively: *ln RD* =-0.39 *ln n* -0.16 ($r^2$ =0.99) and *ln RD* =-0.36 *ln n* -0.02($r^2$=0.99 ); The fitted equations for the corresponding shuffled controls are: for ATG *ln RD* =-0.5 *ln n* -0.02($r^2$=0.99) and for CAT *ln RD* =-0.49 *ln n* - 0.5($r^2$=0.99).

In Fig. 8: the aggregation analysis for the triplet ATG obtained from the 1st and 2nd halves of the ORF-OOO database is shown. For the 1st half of the sequence the distance series of ATG yields an slope $m=-0.31\ (H=0.69; D=1.31; r_1=0.3)$ and for the 2nd half the value of the slope is $m=-0.35\ (H=0.64; D=1.35; r_1=0.23)$. These values indicate long-range correlations dominated by a single scaling exponent. The slope estimate of the complementary triplet CAT in the 1st half of the ORF-OOO sequence ($m=-0.32\ (H=0.68; D=1.32; r_1=0.28)$) is similar to the one obtained for ATG in the 2nd half, whereas for the 2nd half the value of the slope for CAT ($m = -0.35$ ($H = 0.65; D = 1.35; r_1 = 0.23$)) is similar to that of ATG in the 1st half; the corresponding values for the shuffled controls of ATG and CAT indicated that they are random (see caption of Fig. 7). Note that these estimates are essentially the same to those obtained for the intact and intron-containing sequences for each half of the chromosome (Fig. 4).

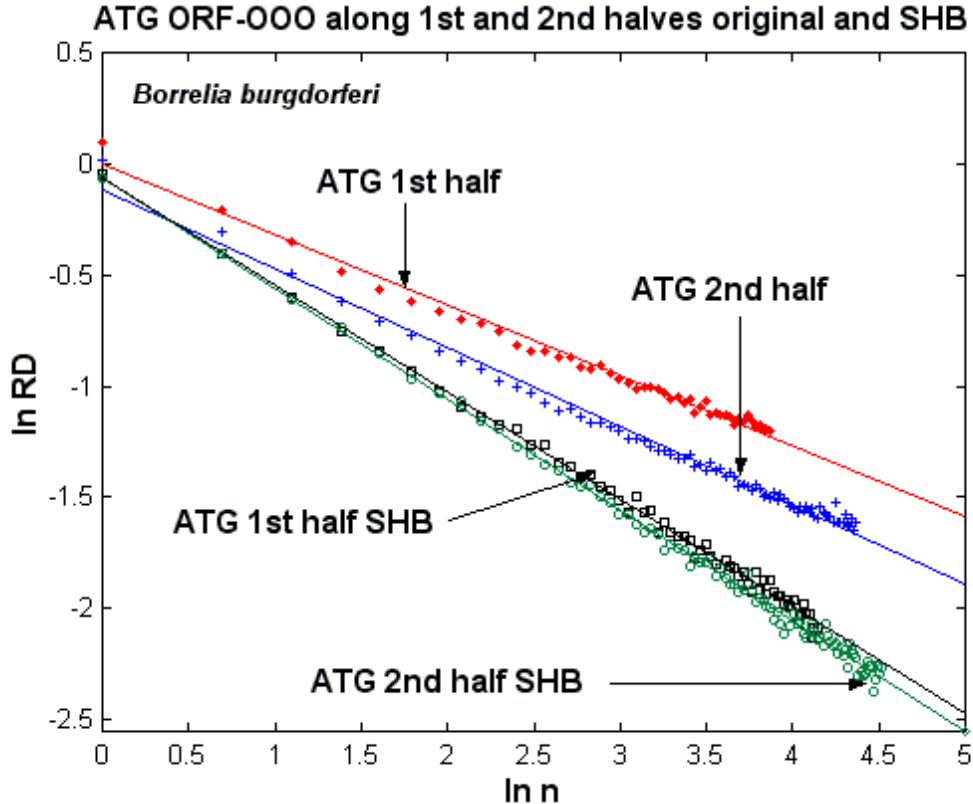

**FIG. 8:** Aggregation analysis of the triplet ATG by halves in the ORF-OOO sequence. For the 1st half the fitted equation is: $\ln RD = -0.31 \ln n - 0.003$ ($r^2 = 0.98$); for the 2nd half the fitted equation is: $\ln RD = -0.35 \ln n - 0.11$ ($r^2 = 0.98$). The fitted equations for the corresponding shuffled controls at the level of bases are for the 1st and 2nd halves, respectively: $\ln RD = -0.48 \ln n - 0.06$ ($r^2 = 0.99$) and $\ln RD = -0.49 \ln n - 0.06$ ($r^2 = 0.99$); For the triplet CAT in the 1st half the fitted equation is: $\ln RD = -0.32 \ln n + 0.04$ ($r^2 = 0.99$); for the 2nd half the fitted equation is: $\ln RD = -0.35 \ln n - 0.11$ ($r^2 = 0.98$) (not shown). The fitted equations for the corresponding shuffled controls at the level of bases are for the 1st and 2nd halves, respectively: $\ln RD = -0.49 \ln n - 0.07$ ($r^2 = 0.99$) and $\ln RD = -0.49 \ln n - 0.03$ ($r^2 = 0.99$).

The CPPs for most triplets and their complementaries in the ORF-OOO sequence of *B. burgdorferi* form a rhombus as is also the case of the complete genome [17]. Here we examine the distance series of triplets along protein-coding regions assembled in the ORF-OOO sequence. Note that the distances at which ATG (or CAT) is encountered along the 1$^{st}$ half (or in the 2$^{nd}$ half) are in average larger than the corresponding distances in the 2$^{nd}$ half (or in the 1$^{st}$ half) of the chromosome (see Figs 9). Thus the distance series for the triplets ATG and CAT are stationary for each half of the chromosome but nonstationary along the whole chromosome. The same pattern holds for almost any triplet of the ORF-OOO sequence and for the whole intact chromosome.

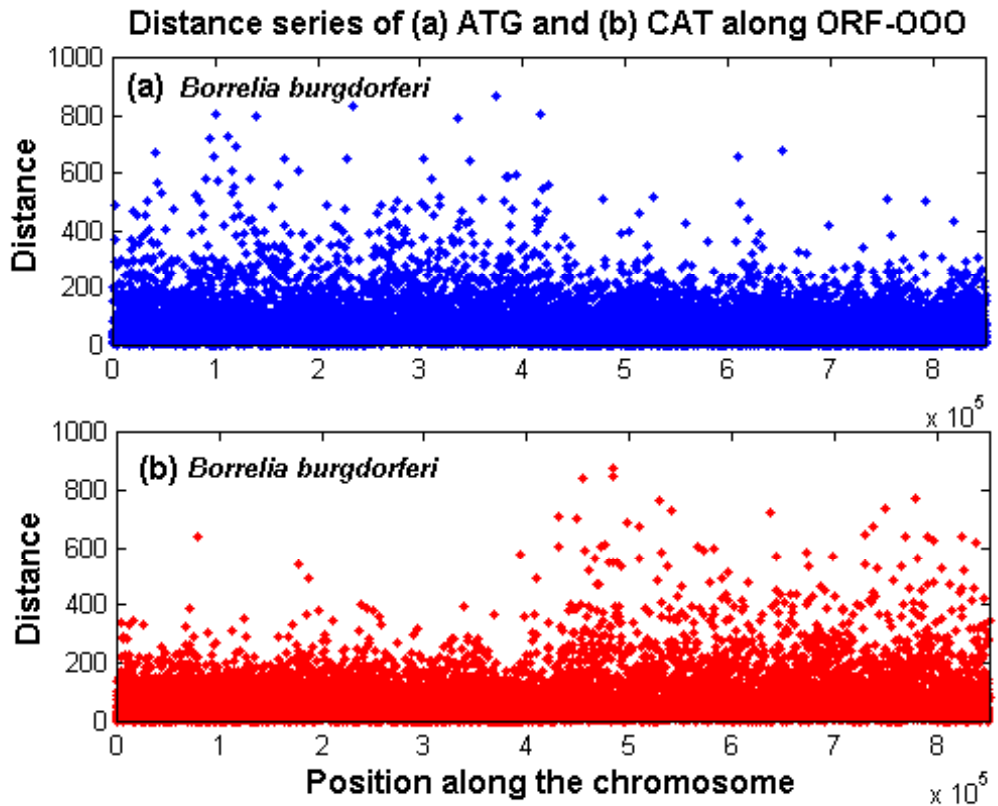

**FIG. 9:** The corresponding distance series of **(a)** ATG and **(b)** CAT along the ORF-OOO sequence of *B. burgdorferi*.

The aggregation analysis for ATG in both the leading and lagging strands also resulted in monofractal series with long-range correlations (for the leading strand: $m = -0.39$ ($H = 0.6; D = 1.39; r_1 = 0.16$); and for the lagging strand: $m = -0.37$ ($H = 0.63; D = 1.37; r_1 = 0.19$) (not shown)).

The aggregation analysis of ATG along the whole ORF-OOO sequence for the original, SWB, and SWT were fitted to Equation (16) indicating that all of them follow an IPL with log-periodic modulation (not shown). The fitted equation for ATG in the ORFs-OOO sequence is:

$$RD^{(n)} = \frac{0.37}{n^{0.36}} e^{0.69\cos[0.56\ln n]}.$$

The fitted equation for the distance series of ATG in the ORF-OOO-SWB sequence is:

$$RD^{(n)} = \frac{0.22}{n^{0.5}} e^{0.98\cos[0.51\ln n]},$$

and for ATG in the ORF-OOO-SWT sequence is:

$$RD^{(n)} = \frac{0.22}{n^{0.5}} e^{0.93\cos[0.56\ln n]}.$$

Note that the scaling exponents for both shuffled controls (SWB and SWT) have a value of: $\hat{\alpha}_{RG} = 0.5$ (i.e. they are uncorrelated) whereas for the original sequence $\hat{\alpha}_{RG} = -0.36$ (i.e. they display long-range correlations).

The log-periodic modulation is still observed by both shuffled controls, SWB and SWT. This is consistent with the fact that their CPPs also generate rhombi. The value of the parameter $b$, which is a measure of the period of oscillation, is of the order of 1.09 and it is the same for the 3 curves.

## 4. Discussion

In this work the scaling properties of DNA sequences taking into account the IBS of the chromosome of *Borrelia burgdorferi* have been examined.
The DNA fluctuations of the distance series of the n-tuples are not random, like Brownian motion, nor are they the result of processes with short term correlations. Instead, the IPL form of the aggregated relative dispersion reveals that the DNA distance series of single bases and triplets at any position is influenced by fluctuations that occurred hundreds or thousands of bases apart. This behavior is a consequence of the self-similar fractal nature of the distance series of DNA.

The distance series of a single nucleotide in each chromosomal half follows a random monofractal but up to a certain distance, beyond which the overall scaling behavior continues to be random but slightly edited by a log-periodic modulation. A similar behavior is obtained for its complementary base in the other half of the chromosome. This result was obtained for each intact (intron-containing) half of the chromosome and the same pattern is observed for the intronless ORF-OOO sequence (not shown).

In contrast, the distance series of triplets in each half of the chromosome or from each half of the ORF-OOO sequence display a monofractal behavior with long-range-correlations which are statistically similar to the corresponding complementary triplet in the other half of the chromosome.

The statistical behaviour of di-nucleotides in each half of the chromosome is intermediate between those of single nucleotides and those of triplets, i.e., they show weak long-range correlations but they are also slightly tuned by log-periodic factors (not shown).

When exclusively protein-coding regions (ORF-CW, ORF-OOO, whole leading and lagging strands) are considered, the resulting distance series of triplets show a monofractal behavior with long-range correlations. If the whole chromosome or the complete ORF-OOO sequence are under study the distance series of the triplets are nonstationary and they become multifractals, so that they are not only long-ranged correlated but they are also modulated by log-periodic factors in agreement with the IBS. Thereby, intergenic and RNA sequences contributed little to both IBS and the scaling properties. This agreed with their low representation (6%) in the chromosome. Similar analysis of other bacterial chromosomes demonstrated similar scaling properties.

In order to study the statistical properties of DNA sequences it is necessary to take into account the IBS exhibited by most Eubacterial chromosomes. This means that it suffices to study the statistical properties of one half of the chromosome. Most bacterial chromosomes exhibit IBS that account all at once for base compositional asymmetry in the chromosome as a whole, including Chargaff's second parity rule [37], the presence of a GC-skew, and bi-directional replication.

The finding of long-range correlations in triplets (or codons) should not be regarded as a surprise since a single codon can be envisaged as the natural biological quantum of information that encodes a single aminoacid.

Higher-order n-tuples may also display IBS [17] and have a monofractal behavior in each half of the chromosome. Such is the case for T-rich octamers including the *ori*-biased [38] TTGTTTTT sequence in *B. burgdorferi*, and the G-rich GCTGGTGG Chi octamer, the *ori*-based signal with a role in recombination associated to chromosomal replication in *Escherichia coli* [39,40].

In this work a RG approach has been used to obtain an expression for the aggregated relative dispersion that is periodic in *ln n* with a fundamental period determined by the scaling parameter *ln b*. The relative dispersion technique, while it is not new, does highlight the modulation of the IPL of single bases in each half of the chromosome allowing to distinguish between the random controls and the actual biological sequences. Other techniques, that smooth the data or that substract the background random noise, suppress these effects (e.g. [41]). In addition it is straightforward and because it does not involve higher order moments of the distance series, is readily accessible.

It is also noteworthy to mention that we obtained the same scaling properties of the DNA sequences by using two independent techniques, namely, the DFA and the relative dispersion. In fact the slope $\alpha_{DFA}$ is effectively equal to the Hurst exponent particularly for monofractal series. When more than one scaling behavior is shown the different scaling behaviors obtained by DFA are related to nonstationarity and hence to multifractality. This is also captured by the renormalization group approach.

The log-periodic modulation is unlikely to represent a random variation, since it is apparent in multiple plots, using data from all single nucleotides and triplets from 10 different Eubacteria that were analyzed in ref [17] (not shown). Rather this modulation is to be associated to the nonstationarity of the distance series and in the case of triplets to IBS when the whole chromosome is considered.

In the literature it often appears that one has only two choices, either a process is a monofractal or it is multifractal. The latter choice, applied to a time series, would imply that the fractal dimension changes over time, ultimately leading to a distribution of fractal dimensions [33]. That is not the situation here, however. The aggregated relative dispersion given by Equation (13) indicates that the process has a preferential scale length, $b$, in addition to the monofractal behavior determined by the IPL index, $\alpha_{RG}$. Thus, there is the interleaving of two mechanisms, one that is scale-free and produces the monofractal, the other has equal weighting on a logarithmic scale and is sufficiently slow as to not disrupt the much faster fractal behavior.

We can see that the scales in the DNA distance series are tied together in two distinct ways. The IPL, or monofractal, as obtained for the chromosomal halves, and the multiplicative time scale giving rise to the log-periodicity. The tying together of the long and the short distance scales may be necessary in order to adaptively regulate the complex DNA sequences in a changing environment. The log-periodic modulation of the IPL is a consequence of the correlation function satisfying a RG relation and having a complex fractal dimension [30]. Sornette argues that the log-periodicity is a result of what he defines as Discrete Scale Invariance (DSI), that is also a consequence of RG properties of the system [30].

The search of long-range correlations in DNA sequences is equivalent to seeking the evolutionary processes behind a power-law distribution. Multiscale properties may arise from the balance between the competition between the tendencies to be ordered (IBS) and disordered (mutations). The fact that DNA sequences at the level of bases contain random components is consistent with the known fact that two of the main ingredients of molecular evolution are mutations and genetic drift which are both random processes. Once natural selection selected for an inverse bilateral symmetry of bacterial chromosomes the two-above random phenomena have operated under the constraints imposed by this type of symmetry.